\documentclass[12pt]{article}

\input epsf
\def\beq{\begin{eqnarray}}
\def\eeq{\end{eqnarray}}
\def\mpl{M_{\rm Pl}}

\def\t{\tilde}
\def\tg{\tilde g}

\def\bga{\bar \gamma}
\def\d{\delta}

\def\lsim{\mathrel{\rlap{\lower3pt\hbox{\hskip0pt$\sim$}}
     \raise1pt\hbox{$<$}}}         %less than or approx. symbol
\def\gsim{\mathrel{\rlap{\lower4pt\hbox{\hskip1pt$\sim$}}
     \raise1pt\hbox{$>$}}}         %greater than or approx. symbol

\usepackage{amsmath}
\usepackage{amsfonts}

\begin{document}

\begin{titlepage}

\begin{flushright}
{NYU-TH-09/08/10}
\end{flushright}
\vskip 0.9cm

\centerline{\Large \bf General Relativity With An Auxiliary Dimension}

\vskip 0.7cm
\centerline{\large Gregory Gabadadze}
\vskip 0.3cm
\centerline{\em Center for Cosmology and Particle Physics}
\centerline{\em Department of Physics, New York University, New York, 
NY, 10003, USA}

\vskip 1.9cm

\begin{abstract}

I consider an extension of General Relativity by 
an auxiliary  non-dynamical dimension that enables our space-time to 
acquire an  extrinsic curvature.  Obtained gravitational equations,  
without or with a cosmological  constant,   
have a selfaccelerated solution  that is independent of the value of 
the cosmological constant, and can describe  the cosmic speedup of 
the Universe as a geometric effect.
Background evolution of  the  selfaccelerated solution 
is identical to that of ordinary de Sitter space.
I show that linear perturbations on this solution 
describe either a massless  graviton, or a massive graviton 
and a scalar,  which are  free of ghosts and tachyons 
for certain choices of boundary conditions.
The obtained linearized expressions suggest that nonlinear 
interactions  should,  for certain boundary conditions, 
be strongly coupled, although this issue is not studied here. 
The full nonlinear Hamiltonian of the theory 
is shown to be  positive for the selfaccelerated solution,  
while in general,  it reduces to surface terms in our and  
auxiliary dimensions.

\end{abstract}

\end{titlepage}

\subsection*{1. Extension of General Relativity}

One simple  way to parametrize the cosmic acceleration \cite {Acc}
is to introduce in the Lagrangian of General relativity (GR) 
the cosmological constant  $\Lambda\sim (10^{-33}~eV)^2$. 
This is not quite satisfactory however, since the parameter 
$\Lambda$ receives contributions from various scales of particle physics 
each of which is many orders of magnitude greater than  
$(10^{-33}~eV)^2$. Without an  underlying principle, cancellation
between these contributions down to  $(10^{-33}~eV)^2$ seems 
conceptually unlikely and technically  unnatural  
\cite {Weinberg}.  

Here we consider an extension of GR, such that  for an arbitrary 
but given value of the parameter  $\Lambda$,  there exists a 
solution -- requiring adjustment of  certain boundary terms -- 
that is independent of $\Lambda$.  Furthermore, the observed cosmic 
acceleration will be due to a new parameter $m$ with the 
dimensionality of mass,  appearing in the 
extended GR Lagrangian.  This parameter does not receive contributions 
from the  particle physics; its value can be set to  
$m\sim 10^{-33}~eV $. The present approach does not explain the smallness 
of $m$;  instead it  gives  a technically 
natural way of describing cosmic acceleration, with  potential 
observational predictions that 
differ form those of GR with the  cosmological constant.
The present approach does  not solve the cosmological constant problem either, 
but instead it reduces the problem to the choice of the 
boundary conditions in the classical gravitational equations, 
with everything else being quantized (more on this in section 2.) 

\vspace{0.1in}

The gravitational  field will be described by an extended  
metric tensor $\t g_{\mu\nu}(x,u)$, with $\mu,\nu=0,1,2,3,$  
which is labeled by a continuous dimensionless parameter $u$. 
The extended  metric  varies as  
$\t g^\prime_{\mu\nu}(x^\prime,u)= \omega _\mu^\alpha (x) 
\omega_\nu^\beta (x) \t g_{\alpha \beta}(x,u) $, 
under the general coordinate transformations 
$x^{\prime \mu} = [\omega^{-1}(x)]^{\mu}_\nu x^\nu$.
This leaves the extended 
interval $ds_u^2 \equiv \t g_{\mu\nu}(x,u)dx^\mu dx^\nu$ \
invariant. However, the matter fields do not depend on $u$; they will 
only couple to the metric tensor
\beq
g_{\mu\nu}(x)\equiv \t g_{\mu\nu}(x,u=0)\,,
\label{ind}
\eeq
with the relevant invariant interval being 
$ds^2 = g_{\mu\nu}(x)dx^\mu dx^\nu$.

Consider the Lagrangian density for the gravitational field
(we use the conventions of \cite {Wald} and 
also set $\mpl^2 =(8\pi G_N)^{-1}=2$, unless stated otherwise)
\beq
{\cal L}= \sqrt{g} R \pm {m^2}\int_{-1}^{+1}du \sqrt{\t g}
\left (  k_{\mu\nu}^2 - k^2 \right )\,,
\label{egr}
\eeq
where $R$ is the Ricci scalar of the metric $g_{\mu\nu}(x)$, while
$k_{\mu\nu}\equiv{1\over 2}\partial_u \t g_{\mu\nu}$,
$k \equiv \t g^{\mu\nu} k_{\mu\nu} $;  all  
indexes in the Einstein-Hilbert term in (\ref {egr}) are raised 
by $g^{\mu\nu} $,  while those in the second term in  (\ref {egr})
by $\t g^{\mu\nu}$. The value of $k_{\mu\nu}$  measures an extrinsic 
curvature of a  $(3+1)$-dimensional constant-$u$  surface in certain  
coordinates in the ``$x-u$ space-time''. The Lagrangian density 
(\ref {egr}) is covariant in $(3+1)$-dimensions.

We impose the ${\bf Z_2}$ symmetry 
on the fields $ \t g_{\mu\nu}(x,u) = \t g_{\mu\nu}(x,- u)$
across the hypersurface $u=0$. Then, it is enough to 
consider the interval $[0, 1]$ for the variable $u$. 
Note that the ``$u$-dimension'' is not dynamical since  
fields have no ordinary derivative terms there. 
Moreover,  integration boundaries in $u$ may in general  take 
any finite value, which can be reduced back to the 
interval   $[-1, 1]$ by an appropriate rescaling 
of $u$ and the parameter $m$, before specifying boundary conditions 
that could be sensitive to such rescaling.

We refer to the surfaces  $u=0,\pm 1$  as fixed  
boundaries.  Eq. (\ref {ind}) imposes one boundary 
condition on the $u$-dependence of the extended metric.
This is not enough to determine 
completely the  $u$-dependence of $ \t g_{\mu\nu}(x,u) $, 
the second boundary condition should also be 
specified.  For this  one can either impose the Neumann-type  
or Dirichlet-type condition at the either boundaries. 
For now we keep this condition unspecified and find 
various solutions that correspond to 
different choices of the second boundary condition.

\vspace{0.5cm}

\subsection*{2. Equations, Solutions and Boundary Terms}

Let us start with the  action of gravity plus ``everything else'':
\beq
S= \int d^4x  \sqrt{g}  \left \{ R 
+ {\cal L}(\Psi, g) \right \}  \pm {m^2} 
\int d^4x \int_{-1}^{+1}du  \sqrt{\t g} 
\left (  k_{\mu\nu}^2 - k^2 \right )\,.
\label{egrS}
\eeq
Here ${\cal L}(\Psi, g)$ is the non-gravitational Lagrangian 
the fields in which  couple  universally to the metric tensor  
$ g_{\mu\nu}(x)$, hence preserving the equivalence principle.  
We'll be looking at very low-energy  phenomena (as compared to  
Planck's scale)  and thus regard  $ \t g_{\mu\nu} $ as an effective 
classical field describing large distance gravitational physics; thus, 
the gravitational part of the  action  will not be quantized 
(it can be regarded as the 1PI effective action in which all the quantum 
loop effects are encoded in the coefficients of various terms). All the other 
interactions encoded in ${\cal L}(\Psi, g)$ will be quantized.

The Lagrangian  ${\cal L}(\Psi, g)$
will  contain in general the cosmological constant generated 
by particle physics.  As noted earlier, it receives contributions from 
the scales of electromagnetic, strong and weak interactions; 
we denoted it by $\Lambda_{fund}$.   Furthermore, 
quantum fluctuations of  the  non-gravitational fields 
in ${\cal L}(\Psi, g)$ will  generate higher dimensional 
gravitational operators, such as $R^2, R^2_{\mu\nu}$, etc.,
which all are functions of $g_{\mu\nu}$ 
and are suppressed by the Planck's scale.  Importantly, non of these terms, 
that are significant in the UV,  can change the effects  of the second 
term in (\ref {egrS}) which switches on in the IR. Moreover, the second 
term in (\ref {egrS}) does not get  renormalized by the quantum loops of 
particle physics, since the particles couple only to  $ g_{\mu\nu}(x)$ and 
cannot give rise to operators made of $ \t g_{\mu\nu} $. This can also 
be seen from the 5D representation of the model given in section 5. There, 
the matter fields localized on the brane cannot renormalize the bulk 
terms because of geometric separation in extra dimension; the bulk terms 
stay unchanged, as long as gravity is considered to be a  classical 
field theory with the effective 1PI action.  

\vspace{0.1in}

The equations of motion obtained by varying the action (\ref {egrS}) 
$\t \delta S$ with the fixed boundary conditions 
in the $u-space$, $\t \delta \t g_{\mu\nu}(x,u)|_{u=\pm 1}=0$
(amended by the condition $\t \delta \t g_{\mu\nu}(x,u)|_{x=boundary,u=0}=0$ 
when the boundary is present in the $x$-space, in which case 
the Gibbons-Hawking (GH) boundary term \cite {GH} should also be 
introduced in the action) gives the following two equations 
for $u=0^+$ and $0<u\leq 1$ respectively:
\beq
G_{\mu\nu} \pm   2 m^2 \left (k_{\mu\nu} - g_{\mu\nu} k \right ) = 
T_{\mu\nu}/2\,, 
\label{junction}
\eeq
and 
\beq
\partial_u 
\left [ \sqrt{\tg} \left (k \tg^{\mu\nu} -
k^{\mu\nu} \right ) \right ] =  {1\over 2} \tg^{\mu\nu }\sqrt{\t g} 
\left ( k^2  -   k_{\alpha\beta}^2   \right ) + 2 
\sqrt{\t g} \left ( k^{\mu\rho} k_{\rho}^{\nu} -k^{\mu\nu}k \right )\,. 
\label{bulk}
\eeq
Note that the right hand side (rhs) of  Eq. (\ref {bulk}) is traceless. 

Furthermore, equation (\ref {junction}) 
combined with the Bianchi identities
implies that:
\beq
D^\mu k_{\mu\nu} = D_\nu k\,,
\label{Con} 
\eeq
where $D_\mu$ denotes  the covariant derivative compatible with 
the metric. Eq. (\ref {Con})  should  automatically be 
satisfied by any solution of (\ref {junction}).
Note that Eqs. (\ref {junction}) and   (\ref {Con}) are similar
to those of the DGP model \cite {DGP} written in the 5D ADM \cite {ADM} 
form (see, e.g., \cite {CedricMourad}).  However, 
there are two significant differences: (a) What is the $\{55\}$ 
equation in  DGP is absent here; (b) In Eq. (\ref {bulk})
there are no derivatives w.r.t. space-time coordinates, 
and thus it significantly differs  from its DGP counterpart 
(what is the bulk $\{\mu\nu\}$ equation).

Equation (\ref {bulk}) determines the evolution of the
metric $\tg_{\mu\nu}$ in the $u$-direction. This is a second order equation.
One boundary conditions for it is specified by (\ref {ind}); pending  the 
second boundary condition  we find different dependence of the metric on 
$u$. The latter  sets the value of the extrinsic curvature
at $u=0^+$, which by its turn determines 4D geometry via Eq. 
(\ref {junction}).

\vspace{0.1in}

We turn now to concrete solutions.   In the absence of any matter 
stress-tensor or cosmological constant ($T_{\mu\nu}=0$) 
the above system of equations has  the Minkowski solution 
$\t g_{\mu\nu}(x,u)=\eta_{\mu\nu} \equiv {\rm diag}\{-1,1,1,1\}_{\mu\nu}$.
In general,  for the class of extended metrics which are independent of 
$u$, the theory at hand reduces to GR. This would 
correspond to the choice of the boundary condition 
$\partial_u \tg_{\mu\nu}|_{u=0}=0$, in addition to (\ref {ind}). 
Thus, for $\t g_{\mu\nu}(x,u)=\eta_{\mu\nu}+h_{\mu\nu}(x)$
the fluctuations of the extended metric above the Minkowski 
solution describe a massless graviton.

There exists a choice of the boundary conditions 
for which the linearized fluctuations describe a Minkowski space 
massive graviton; for instance,  the Lagrangian (\ref {egr}) with 
the minus sign in front of the second term describes the 
Pauli-Fierz massive graviton of $ (\rm mass)^2= 2m^2$,  
with  $\t g_{\mu\nu}(x,u)=\eta_{\mu\nu}+ (1- |u|)h_{\mu\nu}(x)$ being 
a linearized solution selected  by imposing the second 
boundary condition in the Dirichlet form: 
$\tg_{\mu\nu}(x,u)|_{u=1}=\eta_{\mu\nu}$.

Hence, the theory (\ref {egr}) endowed with the appropriate boundary 
conditions  gives a nonlinear completion of massive gravity. 
Remarkably, the Hamiltonian of this theory does not  suffer from the 
problem found in Ref. \cite {Deser} in 4D massive gravity, 
as it will be shown in Section 4. 
Since Minkowski space is not a subject of a primary interest here, we 
will not elaborate on this branch of solutions further.

\vspace{0.1in}

Consider now a factorized  expression for the extended metric  
\beq
\t g_{\mu\nu}(x,u)= a(u) g_{\mu\nu}(x)\,. 
\label{a}
\eeq
The rhs  of Eq. (\ref {bulk}) is identically zero for 
(\ref {a}), and Eq.  (\ref {bulk}) reduces to $\partial_u^2a=0$.  
Hence, for $u\ge 0$ we have $a(u) =c_0 +c_1  u$, where $c_0,c_1 $ are 
integration constants. The boundary condition (\ref {ind}), 
and (\ref {a}) define the value of $c_0=1$, while $c_1$ has to be 
fixed by the second boundary condition. Below we consider various  
solutions that differ from each other by the choice of that 
boundary condition.

For the second boundary condition specified in the following form 
\beq
\partial_u \tg_{\mu\nu}|_{u=0^+}= \mp g_{\mu\nu}(x)\,, 
\label{bc1}
\eeq
it is straightforward and not tedious  to check that the system  of 
equations (\ref {junction}),(\ref {bulk})  admits 
a selfaccelerated solution:
\beq
\t g^{cl}_{\mu\nu}(x,u) \equiv (1\mp  |u|){\bar \gamma}_{\mu\nu}(x)\,,~~~~~
R(\bga) = 12 m^2\,.
\label{sa}
\eeq
Here, ${\bar \gamma}_{\mu\nu}(x)$ denotes the 4D de Sitter 
metric with the Hubble parameter $H$ equal to $m$. This solution can describe 
the cosmic acceleration of the Universe, with the acceleration 
being due entirely to a geometric effect.
In that regard, the growing solution in (\ref {sa}) is similar 
to the selfaccelerated solution \cite {Cedric,DDG} on the DGP model,
while the decaying solution to that of Refs. \cite {GGCargese}.

For the decaying solution in (\ref {sa})
the extended metric $\tg_{\mu\nu}$  vanishes at  the boundaries  $u=\pm1$,
while the inverse of $\tg_{\mu\nu}$ is singular,  giving rise to 
a singularity  of the extended Ricci tensor $\t R$ made of $\tg_{\mu\nu}$.  
However,  since  the ``$u$-dimension'' is  nondynamical, and all the matter 
and their  interactions are located at  $u=0$,  the extended Ricci tensor 
$\t R$  evaluated at $u=\pm1$ should not have a  particular significance.
Moreover, this singularity is easily avoidable by changing 
in (\ref {egr}) the integration 
interval for $u$ from $[0,1]$ to $[0,b]$, where $b<1$ is some 
positive number. This would not change the equations (\ref {junction}) and 
(\ref {bulk})  and the solution (\ref {sa}), but 
for any $b\neq 1$ one would need to add to the 
Lagrangian  (\ref {egr}) a surface term. The latter 
would  guarantee that the effective Lagrangian obtained by  
integrating over  the  $u$-direction (i.e., by first substituting the 
metric (\ref {a}) into the action and then varying it w.r.t. the metric 
$g$)  gives the result consistent with  the solution (\ref {sa}) 
obtained from the equations of motion (\ref {junction}) and (\ref {bulk}).

The Lagrangian with the surface terms for general $b$,  which gives the 
selfaccelerated solutions (\ref {sa}),  reads  as follows:
\beq
{\cal L}_b= \sqrt{g} R \pm {m^2}\int_{-b}^{+b}du \sqrt{\t g}
\left (  k_{\mu\nu}^2 - k^2 \right )+ C^{(\pm)}_b m^2  \left (  
\sqrt{\t g}|_{u=b}+\sqrt{\t g}|_{u=-b}\right )\,,
\label{egrb}
\eeq
where $C^{(\pm)}_b \equiv - 3/(1 \mp b)$.

For the growing solution with the positive sign   in (\ref {sa})
the above singularity is absent, however,  
even when  $b=1$ is chosen,   for this solution one has to add to the 
Lagrangian the surface term given in (\ref {egrb}) in order for the effective  
Lagrangian (obtained by  integrating  out the $u$-direction)  
to give the result consistent with the solution (\ref {sa}) that was 
derived  from the equations of motion (\ref {junction}) and (\ref {bulk}).
Moreover, for the growing solution in (\ref {sa}) 
the  surface term  will be crucial 
for calculation of its energy in Section 4.

Although we have constructed these surface terms  via ``inverse engineering''
staring with the desired solutions, the straightforward 
statement is the following one:  for the given boundary conditions and 
specified surface terms there are unique selfaccelerated solutions 
corresponding to the two sign choices in  (\ref {egr}). 

\vspace{0.1in}

One could of course modify the second boundary condition
(\ref {bc1}) in various ways  and obtain different solutions,
to some of which we're turning now. 

For a nonzero homogeneous and isotropic 
stress-tensor there exists a solution for which the  
extended metric reads $\t g_{\mu\nu}(x,u) 
\equiv (1\mp \zeta |u|)\gamma^{FRW}_{\mu\nu}(x)$,
and  the modified Friedmann equation in the 
standard notations takes the form 
\beq
H^2 -\zeta m^2  + {{\kappa}\over a^2} = {8\pi G_N\over 3} \rho \,,  
\label{Fr}
\eeq
where ${\kappa}=\pm1,0$  labels the 3D spatial curvature, 
and $\zeta$ is an arbitrary integration constant that could be 
fixed only  after imposing the boundary condition 
for e.g.,  $\t g_{\mu\nu}(x,u)|_{u=\pm 1}$, or 
for $\partial_u \t g_{\mu\nu}(x,u)|_{u=0,\pm1}$. 

If the stress-tensor contains the cosmological constant 
($8\pi G_N T_{\mu\nu}= \Lambda_{fund} g_{\mu\nu}$) the value of 
$\zeta$ can be chosen to cancel its contribution down 
to zero.  This can be combined with the selfaccelerated solution 
obtained above. For instance, consider the Lagrangian with the 
cosmological constant  and the choice of the plus sign in front 
of the second term in (\ref {egr}): 
\beq
{\cal L}_b= \sqrt{g} (R -2 \Lambda_{fund}) + {m^2}\int_{-1}^{+1}
du \sqrt{\t g}
\left (  k_{\mu\nu}^2 - k^2 \right )+  C_\zeta m^2  
\left ( \sqrt{\t g}|_{u=b} + \sqrt{\t g}|_{u=-b} 
\right ).
\label{egrb1}
\eeq
The corresponding equations (\ref {junction}) and 
(\ref {bulk}) have a consistent solution:
\beq
\t g^{cl}_{\mu\nu}(x,u) \equiv 
(1+ \zeta |u| - |u|) {\bar \gamma}_{\mu\nu}(x)\,,~~~~~
R(\bga) = 12 m^2\,,
\label{saLambda}
\eeq
if $C_\zeta =  3 (\zeta-1) /\zeta $, where 
$\zeta \equiv \Lambda_{fund}/3m^2 \gg 1$.

The result of this discussion is the following:
for an arbitrary value of the cosmological constant 
generated by particle physics  $\Lambda_{fund}$,  
one can choose the boundary conditions and surface 
term in (\ref {egrb1}) such  that the background solution describes an  
accelerated universe  with the Hubble parameter that is 
independent of  $\Lambda_{fund}$,  but instead 
is defined by the UV insensitive 
new mass scale $m$,  introduced in 
the Lagrangian (\ref {egr}), or  (\ref {egrb1}).

This scheme does not provide a dynamical mechanism for solving the 
cosmological constant problem, as one has to adjust the boundary terms 
and conditions appropriately to get rid of $\Lambda_{fund}$.
However, it has an advantage over GR  in the following respect:
GR,  as well as  the present model,  at the classical level 
can entirely be specified by their equations of motion,   
without any reference to the action. The GR equations with the  
cosmological constant have no other  solutions  but the (A)dS solutions 
with curvature set by the value of the cosmological constant.
In contrast with this,  the equations of motion of the present theory 
with the cosmological constant  do  have solutions with curvatures 
that are not related  to the cosmological constant.  
The above properties of the equations  
make no reference to the boundary terms. The latter 
come into the play only when the action functional is invoked.

Hence, as long as gravity is treated classically while all the other 
interactions are quantized, the present approach reduces the cosmological 
constant  problem to the choice of the boundary 
conditions in the classical gravitational equations.

The fact that $\Lambda_{fund}$ can be removed by means of  
the boundary conditions which specify the otherwise arbitrary 
integration constant, is  somewhat similar  to what happens in 
the unimodular gravity \cite {Uni1,Uni2} where the cosmological 
constant can be fixed by superselection rules. 
However, a  distinction  between the two approaches 
is that the perturbations in the present 
case can be different from those of the unimodular 
gravity  which are  identical to the GR perturbations.

In the  context of  inflationary cosmology, the present 
method would  remove a constant piece from the inflationary potential, 
while retaining all the positive aspects of the slow-roll 
inflationary paradigm (note a similarity in this 
with Ref. \cite {ADDG}).

\vspace{0.1in}

As mentioned before, the theory (\ref {egr}) contains all the solutions of 
GR:  using the factorized form  (\ref {a}) with $a=1$ 
one would obtain  just Einstein's equations  
for $g_{\mu\nu}$. For the selfaccelerated universe
$a=1\mp |u|$, and  equation (\ref {junction}) for 
$g_{\mu\nu}$ is  the ordinary 
Einstein equation with the cosmological constant equal to $3m^2$. Thus, 
for instance, the dS-Schwarzschild solution of GR is also 
a factorized solution on the selfaccelerated background.
Similar arguments apply to  any other 
solution of the Einstein equations. Furthermore, 
there may well  exist other solutions, e.g., for a static source, 
that  do not have the factorized form (\ref {a}). The latter would be 
selected from the factorized solutions by the boundary conditions.  

Factorized or not,  the spectrum of linear and/or nonlinear 
perturbations about these solutions are determined by Eqs. (\ref {junction}), 
(\ref {bulk}), which themselves may or may not  have a factorized 
form (\ref {a}),  or depending on boundary conditions, 
could differ from the spectrum of GR. One example of this 
is the spectrum of linear perturbations on the selfaccelerated
solution to which we turn in the next section.

\newpage

\vspace{0.1in}

\subsection*{3. Perturbations of the selfaccelerated solution}

We denote the deviation from the background metric as follows: 
\beq
\tg_{\mu\nu}(x,u)= {\t g}^{cl}_{\mu\nu}(x,u)+ \d g_{\mu\nu}(x,u)\,. 
\label{pert}
\eeq
Where, ${\t g}^{cl}$ is defined in (\ref{sa}). It is straightforward 
to derive that 
\beq
k_{\mu\nu}={\bar k}_{\mu\nu}+  \d k_{\mu\nu}\,,~~~ {\bar k}_{\mu\nu}=
\mp {1\over 2} {\bar \gamma}_{\mu\nu}\,,~~~
k ={\bar k}+  \d k\,, ~~~{\bar k}= {\bar g}^{\mu\nu} {\bar k}_{\mu\nu}=
\mp {2\over a}\,,
\label{kdk}
\eeq
where 
\beq
\d k_{\mu\nu}= {1\over 2} \partial_u \d g_{\mu\nu}\,,~~~~~
\d k= {1\over 2a}{\bar \gamma}^{\mu\nu} \partial_u \d g_{\mu\nu}
\pm {1\over 2a^2}{\bar \gamma}^{\mu\nu} \d g_{\mu\nu}\,.
\label{dk}
\eeq
An expansion of Eq. (\ref {junction}) in the linear approximation reads:
\beq
\d G_{\mu\nu} \pm 2m^2 \left (\d k_{\mu\nu} - \d g_{\mu\nu}{\bar k}-
{\bar g}_{\mu\nu}  \d k   \right ) = T_{\mu\nu}/2\,,
\label{djunction}
\eeq
here $T_{\mu\nu}$ on the r.h.s. is the stress-tensor of a probe
source which has nothing to do with the background; 
the variation of the Einstein tensor on the dS space is
\beq
\d G_{\mu \nu}= -{1\over 2} \left ( \square \d g_{\mu\nu} - \nabla_\mu
 \nabla^\alpha \d g_{\alpha \nu} - \nabla_\nu
 \nabla^\alpha \d g_{\alpha \mu} + \nabla_\mu \nabla_\nu \d g^{\alpha}_
{\alpha} \right ) \nonumber \\
-{1\over 2} {\bar \gamma}_{\mu\nu}
\left (\nabla^\alpha \nabla^\beta \d g_{\alpha \beta} -  
\square \d g^{\alpha}_{\alpha}  \right ) 
- 2H^2 \d g_{\mu\nu} +{1\over 2}  H^2 {\bar \gamma}_{\mu\nu} 
\d g^{\alpha}_{\alpha}\,,
\label{dG}
\eeq
where $\nabla$ denotes the  covariant derivative w.r.t. $\bga$.
The constraint (\ref {Con}),  
which in the linearized approximations reads
\beq
\pm \nabla^\mu \d g_{\mu\nu} + \nabla^\mu \partial_u \d g_{\mu\nu} =
\pm \nabla_\nu \d g^{\alpha}_{\alpha} + \nabla_\nu \partial_u 
\d g^{\alpha}_{\alpha}\,,
\label{Conlin}
\eeq  
can be satisfied by the following gauge fixing condition 
\beq
\nabla^\alpha \d g_{\alpha \beta} = \nabla_\beta \d g^{\alpha}_{\alpha}\,.
\label{gProca}
\eeq
Using the  latter in equation (\ref {djunction}), where we also 
substitute $m^2=H^2$, we obtain:
\beq
-{1\over 2} \left 
( \square \d g_{\mu\nu} - \nabla_\mu \nabla_\nu \d g^{\alpha}_
{\alpha} \right ) + 2H^2 \d g_{\mu\nu} - {1\over 2} \bga_{\mu\nu} 
H^2  \d g^{\alpha}_{\alpha} \nonumber \\
\pm H^2 \left ( \partial_u \d g_{\mu\nu} - \bga_{\mu\nu}  
\partial_u \d g^{\alpha}_{\alpha}
\right ) = T_{\mu\nu}/2 \,.
\label{dggauge}
\eeq
Taking  trace of the above equation gives:
\beq
\mp 3 H^2 \partial_u \d g^{\alpha}_{\alpha}= T/2\,.
\label{trace}
\eeq
One  needs to solve  equation  (\ref {bulk}) 
to obtain the $u$-dependence of the perturbations. 
For this one considers variation of its left and right 
hand sides separately at $u>0$:
\beq
\d \left \{ \partial_u 
\left [ \sqrt{\tg} \left ( k \tg^{\mu\nu} -
k^{\mu\nu} \right ) \right ] \right \} = \nonumber \\  \sqrt {\bga} \partial_u 
\left \{ {1\over 2} \bga^{\mu\nu} \bga^{\alpha \beta} 
\partial_u \d g_{\alpha \beta}
\mp {1\over 4a} \bga^{\mu\nu} \bga^{\alpha \beta}  \d g_{\alpha \beta}  
-{1\over 2} \bga^{\mu\alpha} \bga^{\nu \beta} \partial_u \d g_{\alpha \beta}
\pm {1\over a} \bga^{\mu\alpha} \bga^{\nu \beta} \d g_{\alpha \beta} 
\right \}\,.
\label{dlhs}
\eeq
Notice that all the equations presented above 
with the upper sign are equivalent to those with the 
lower sign provided that in the latter the replacement 
$u \to - u $ is made. This will be reflected in our final solutions.

The variation  of the rhs of (\ref {bulk}) equals to:
\beq
\sqrt{\bga}  \left \{   
{1\over a^2} \bga^{\mu\alpha}\bga^{\nu \beta} \d g_{\alpha \beta} \pm 
{1\over a} \bga^{\mu\alpha}\bga^{\nu \beta} \partial_u 
\d g_{\alpha \beta}  - {1\over 4a^2}\bga^{\mu\nu} 
\bga^{\alpha \beta}  \d g_{\alpha \beta} 
\mp {1\over 4a}\bga^{\mu\nu} \bga^{\alpha \beta} \partial_u 
\d g_{\alpha \beta}\, 
\right \}\,.
\label{drhs}
\eeq
Putting Eqs. (\ref {dlhs}) and (\ref {drhs}) together,  
certain cancellations occur, and  we find the final equation:
\beq
\bga^{\mu\nu} \bga^{\alpha \beta} \partial^2_u \d g_{\alpha \beta} =
\bga^{\mu\alpha}\bga^{\nu \beta} \partial^2_u \d g_{\alpha \beta}\,.
\label{dfin}
\eeq
The latter has a solution 
\beq
\d g_{\alpha \beta}= (1+c u) h_{\alpha\beta}(x), 
\label{sol1}
\eeq
where $c$ is an arbitrary constant to be fixed by the boundary 
conditions\footnote{The expression in (\ref {sol1}) is not a most 
general solution of 
(\ref {dfin}), however, it can be selected among all the solutions 
by specifying appropriate boundary conditions (see below).}.
The two  sign  choices considered above will hereafter be 
encoded in the value of $c$. We'll keep this constant 
unspecified till the end of our calculations.

Using the solution (\ref {sol1}) in equation (\ref {dggauge}) we find: 
\beq
-{1\over 2} \left ( \square h_{\mu\nu} - \nabla_\mu \nabla_\nu h^{\alpha}_
{\alpha} \right ) + 2H^2 h_{\mu\nu} - {1\over 2} H^2 
\bga_{\mu\nu}   h^{\alpha}_{\alpha} 
\nonumber \\
+ H^2 \,c\, \left ( h_{\mu\nu} - \bga_{\mu\nu}  h^{\alpha}_{\alpha}
\right ) = T_{\mu\nu}/2 \,,
\label{eq1}
\eeq
with its trace equation
\beq
-3H^2\,c\,h=T/2\,.
\label{trace1}
\eeq
Combining the above  two equations, introducing the 
Lichnerowicz operator which in our case satisfies:
\beq
\Delta_L h_{\mu\nu} = -\square h_{\mu\nu} + 
8H^2  h_{\mu\nu}-2 H^2 \bga_{\mu\nu}
h^{\alpha}_{\alpha}\,,
\label{Lich}
\eeq
and using the standard techniques  (see, \cite {DGI,GIS} for recent 
discussions), we obtain the following expression for the perturbations:
\beq
h_{\mu\nu}= {1\over \Delta_L - 6H^2 +2H^2(c+1) }T_{\mu\nu}  
-{1\over 3} \bga_{\mu\nu} {1 \over -\square - 6H^2 +2H^2(c+1) }T \nonumber \\
+{1\over 6c} \bga_{\mu\nu} {1 \over -\square - 6H^2 +2H^2(c+1) }T +
{\nabla_\mu \nabla_\nu \over 6H^2 c} {1\over  -\square - 6H^2 +2H^2(c+1)}T\,. 
\label{hmunu}
\eeq
The expression for the physical one-particle exchange 
amplitude reads as follows:
\beq
{\cal A} \equiv \int d^4x \sqrt {\bga} T^{\prime \mu\nu} h_{\mu\nu} = 
\int d^4x \sqrt{\bga}  T^{\prime \mu\nu} {1\over \Delta_L - 6H^2 
+2H^2(c+1) }T_{\mu\nu}  \nonumber \\
-  \int d^4x \sqrt{\bga}  \left (  {1\over 3} -
{1\over 6c} \right ) T^\prime {1 \over -\square - 
6H^2 +2H^2(c+1) }T \,,
\label{A}
\eeq
where $ T^{\prime \mu\nu} $ denotes another conserved probe source.
This should be compared with the amplitude for a massless graviton on dS 
space
\beq
{\cal A}_0\equiv \int d^4x \sqrt{\bga} T^{\prime \mu\nu} h_{\mu\nu}=
\int d^4x \sqrt{\bga}\left \{  T^{\prime \mu\nu} {1\over \Delta_L - 6H^2 }
T_{\mu\nu} - {1\over 2} T^\prime {1 \over -\square - 
6H^2  }T  \right \} ,
\label{A0}
\eeq
or with the amplitude for a massive graviton of mass $M$ on dS space
\beq
{\cal A}_M\equiv \int d^4x \sqrt{\bga} T^{\prime \mu\nu} h_{\mu\nu}
=~~~~~~~~~~~~~~~~~~~~~~~~~~~~~~~ \nonumber \\
\int d^4x \sqrt{\bga}\left \{  T^{\prime \mu\nu} {1\over \Delta_L - 6H^2+M^2 }
T_{\mu\nu} - {1\over 3} T^\prime {1 \over -\square - 
6H^2+M^2  }T \right \}.
\label{AM}
\eeq
For  $c=-1$ the amplitude (\ref {A}) is equivalent to that  
of a massless tensor field on dS in GR (\ref {A0}).
The solution for the background plus its 
perturbation  in this case reads as follows:
\beq
\tg_{\mu\nu}= (1\mp |u|) \bga_{\mu\nu}+  (1\mp |u|)h_{\mu\nu}\,,
\label{pert0}
\eeq
where $h_{\mu\nu}$ is given in (\ref {hmunu}). Note that these 
solutions corresponds to choosing (\ref {bc1}) as the second 
boundary condition.

For $c>2$ one gets a massive graviton on the dS  background \cite {Higuchi}
and a massive scalar with the graviton mass $M^2= 2H^2(c+1)$ and the 
scalar mass $M_s^2= 2H^2(c+1) - 6H^2$. Moreover, the scalar couples 
to the stress-tensor with the  $1/c$ suppressed strength  as 
compared with the gravitational coupling. 
The metric for the  solutions takes the form:
\beq
\tg_{\mu\nu}= (1\mp |u|) \bga_{\mu\nu}+  (1\pm  c|u|)h_{\mu\nu}\,.
\label{pertM}
\eeq
This solution corresponds to choosing in addition to (\ref {ind}) 
the following  boundary condition: $\partial_u \tg_{\mu\nu}|_{u=0^+}
= \mp \left ((1+c)\bga_{\mu\nu}-cg_{\mu\nu} \right )$.

For $c=0$ the solution exist only for conformal sources with $T=0$, 
for which  one gets a special massive tensor on dS background with 
enhanced symmetry \cite {DeserW}.

The boundary conditions with values of $c$ other than $c=-1, c=0$ and  
$c\ge 2$ give rise to instabilities: for $c<-1$ one gets a tachyonic 
tensor field 
(which implies that its helicity-0 component is a ghost) and a 
ghost-like scalar; for  $-1<c<0$ one gets a massive tensor and a 
tachyonic scalar ghost; For $0<c<2$ one gets massive  tensor and a 
tachyonic scalar.

Note that the last term in the expression for the field 
(\ref {hmunu})  is singular in the $H=m\to 0$ limit.
This term  does not enter the linearized amplitude,  but as it is 
well known,  such terms typically give rise to strongly coupled behavior
of massive theories  \cite {Arkady,DDGV,AGS,Rat1,Rub}. This is 
a welcome feature in a classical theory as it 
provides a way to overcome the vDVZ discontinuity \cite {vDVZ}, 
as  was first argued by Vainshtein \cite {Arkady}  
(see also \cite {DDGV}, and more recent works
\cite {CedricB}). The magnitude of the strong scale should grow 
with $c$, as it's suggested  by (\ref {hmunu}). More detailed 
questions on its dependence on boundary terms and conditions are left open.
The perturbative results obtained above have a limited 
applicability as the theory is expected to be strongly coupled. Moreover, 
perturbative stability is not a guarantee of a stability of the 
full nonlinear theory, however, it is a first and important step on 
the way to establish whether or not the theory could be viable.

\vspace{0.5cm}

\subsection*{4. Hamiltonian}

In this section we derive the Hamiltonian for the theory (\ref {egr}).
For this we use  the standard ADM decomposition \cite {ADM}:
\beq
 \tg^{00} \equiv -{1\over N^2}, ~~\tg_{0j}\equiv N_j, ~~~
\tg_{ij}\equiv \gamma_{ij}\,,  \nonumber \\
\tg_{00} = -(N^2 - N_i\gamma^{ij}N_j)\,,~~
\tg^{0j} =  {N^j\over N^2},~~\gamma^{ij}=\tg^{ij}+ {N^i N^j\over N^2}\,. 
\label{ADM}
\eeq
After somewhat lengthy  algebra the additional term in the 
Lagrangian (\ref {egr}) can be written as\footnote{Note that in this 
section $\gamma$ refers to the 3D metric, as defined in (\ref {ADM}), 
while  $\bga$ denotes, as before,  the 4D de Sitter metric.}:
\beq
\sqrt{\t g}
\left (  k_{\mu\nu}^2 - k^2 \right )= \sqrt{\gamma}\left 
( N (q_{ij}^2-q^2) - { V^j \gamma_{jk}V^k \over 2N }-
2 q \partial_u N \right )\,,
\label{mgrADM}
\eeq
where all indexes are raised by $\gamma^{ij}$;  
$q_{ij}\equiv {1\over 2}\partial_u \gamma_{ij}=k_{ij}$,
$q\equiv \gamma^{ij}q_{ij}$, and $V^j \equiv \partial_u N^j$.

The expression  in (\ref {mgrADM})
does not contain any time derivatives. Therefore, the canonical momenta
in the extended theory (\ref {egr}) are the same as in GR.
The Hamiltonian density can straightforwardly be calculated
\beq
{\cal H}_u = \sqrt {\gamma} (NR^0 +  N_j R^j)\delta (u)  
 \mp {m^2} \sqrt{\gamma}\left 
( N (q_{ij}^2+ q^2) - { V^j \gamma_{jk}V^k \over 2N }+ 
2 N \partial_u q \right )+ \Sigma\,.
\label{mgrHu}
\eeq
Here $\Sigma$ denotes the surface terms for both, the possible 
spatial boundaries, as well as 
the boundaries in the ``$u$-dimension''
\beq
\Sigma \equiv 2 \nabla_j (  \gamma^{-1/2} N_k \pi^{kj} ) \delta(u) \pm 
2 {m^2} \partial_u \left ( \sqrt{\gamma} q N \right )\,.
\label{S}
\eeq
The first two terms in   (\ref{mgrHu}) 
(the ones that are multiplied by $\delta(u)$)  
are those of GR with  $R^0\equiv - R_{(3)}+ \gamma^{-1}
(\pi_{ij}^2 - {1\over 2} \pi^2 )$, and 
$R^j \equiv -2 \nabla_k (\gamma^{-1/2}\pi^{kj})$, with 
$\pi_{ij}$ being the canonical momenta of GR (see, e.g. \cite {ADM},
\cite {Wald}). 

Since the additional terms in the Lagrangian (\ref {egr}) have no time 
derivatives (\ref {mgrADM}), the primary constraints  of GR 
are preserved;  the conjugate momenta for  the lapse $N$ and shift 
$N_j$ are zero,  $P_N=P_{N_j}=0$.  
Hence, variation of the Hamiltonian  $\t \delta H$ under the variations of 
$\t \delta N$ (such that  $\t \delta N(x, u)|_{u=\pm 1}=0$, and 
vanishing variation at $u=0$ and  $x=\rm boundary$) and 
$\t \delta N_j$ (such that  $\t \delta N_j(x, u)|_{u=\pm 1}=0$
and vanishing variation at $u=0$ at $x=\rm boundary$)\footnote{If the 
boundary is present in the $x$-space the GH boundary term should also 
be introduced.} leads respectively to the following  relations:
\beq
R^0\delta(u) \mp  2 m^2 \sqrt{\gamma} \partial_u q  \mp {m^2}
\sqrt{\gamma} \left ( q_{ij}^2+q^2 + { V^j \gamma_{jk}V^k \over 
2N^2} \right )=0\,,
\label{N} \\\
\sqrt {\gamma} R^j\delta(u)  \mp  {m^2} \gamma^{jk} \partial_u \left [  
\sqrt{\gamma} { \gamma_{ki}V^i \over N} \right ] =0\,.
\label{Nj}
\eeq
Substituting these into the expression for the 
Hamiltonian  (\ref {mgrHu}), one finds that the ``bulk'' terms all 
cancel  and what is left is just the boundary terms:
\beq
H(t) = \int d^3x \int_{-1}^{+1}du\, {\cal H}_u(x) = 
\pm 4m^2 \int d^3x \sqrt{\tg} (\tg^{ij}k_{ij} +\tg^{0i}k_{0i})|^{+1}_0,
\label{mgrH}
\eeq
where we  used the relation 
$
 \sqrt{\gamma} (Nq  + 
  ({N_j\partial_u N^j/ 2N})) = 
\sqrt{\tg} (\tg^{ij}k_{ij} +\tg^{0i}k_{0i})
$, 
and dropped the surface term  that appears in the 
GR Hamiltonian (that is the first term in (\ref {S})).
Note that this is a Hamiltonian that follows from 
the Lagrangian (\ref {egr}). If one adds additional surface terms 
as in (\ref {egrb}),  those terms should simply be 
subtracted from (\ref {mgrH}) 
to get the right Hamiltonian.

For illustration  we calculate the energy  for the selfaccelerated solution
(\ref {sa}) with $a=1-|u|$. The result is positive:
\beq
H(t) = 6 m^2 \int d^3x \sqrt{\bga}\,.
\label{saH}
\eeq
For the selfaccelerated solution with the growing $a(u)$ 
in (\ref {sa}) the calculation of energy gives the same 
result (\ref {saH}) only after inclusion of the boundary term 
given in (\ref {egrb}).

As we  see, the positive semi-definiteness of the Hamiltonian
(\ref {mgrH}),  in which the constraints (or algebraically determined 
relations) were used,    depends on the boundary conditions in the  
$u$-direction.  However, making these boundary terms positive 
semi-definite does not in 
general guarantee absence of instabilities, since the latter can be 
``hidden'' in the  constraint equations.  One example of this is GR with a 
minimally coupled scalar of a negative kinetic term. The GR constraints put 
the Hamiltonian of this system to be zero, however, there are instabilities 
in the theory already at the classical level.

In our case, the above derived results  
can be used to deduce the following important 
observation: In the $m\to 0$ limit the Hamiltonian  (\ref {mgrH}) 
goes to zero. This is in contrast with the $1/m^2$ term in  
the Hamiltonian for 4D massive gravity found by Boulware and Deser 
in \cite {Deser}.  Moreover, the expression (\ref {mgrH}) has no 
singular behavior in  the field fluctuations,  that was found 
in \cite {Deser} as a source of various instabilities in massive gravity.
Hence, even though there is no complete proof of 
the absence of instabilities in the full non-linear theory, 
the absence of the Boulware-Deser singular 
term  in the Hamiltonian is a promising 
step forward.

\vspace{0.1in}

From Eq. (\ref {N}) 
we find two equations for $u=0 $ and $u> 0 $ respectively
\beq
\sqrt {\gamma} R^0|_{u=0}= \pm 2m^2 \sqrt {\gamma} q|^{0^+}_{0^-}\,,~~~~
\sqrt {\gamma}\partial_u q = -{1\over 2}\sqrt {\gamma} 
\left ( q_{ij}^2 + q^2 + { V^j \gamma_{jk}V^k \over 
2N^2} \right )\,,
\label{N2}
\eeq
where the rhs of the last equation is positive semi-definite. 
Similarly,  we obtain from (\ref {Nj}) the following equations 
for $u=0 $ and $u>0$ respectively: 
\beq
\sqrt {\gamma} R^j |_{u=0} = \pm {m^2\over N}\sqrt {\gamma} 
V^j |^{0^+}_{0^-}\,,~~~~
\partial_u \left ( \sqrt{\gamma} {\gamma_{ji}
V^i \over N }     \right )=0\,. 
\label{Nj2}
\eeq
Let us count the degrees of freedom. The variables  
$N|_{u=0}$ and  $ N_j|_{u=0}$ can be fixed by gauge 
transformations. The variables $\partial_u  N|_{u=0}$ and 
$\partial_u  N_j|_{u=0} $ can also be  fixed after choosing  
the boundary conditions,  for instance as  
$N|_{u=\pm 1}$ and  $ N_j|_{u=\pm 1}$,   and using  the 
equations (\ref {N2}) and (\ref {Nj2}).  After fixing the 
boundary conditions what is left undetermined is the 12  
variables $\gamma_{ij}|_{u=0}, \pi_{ij}|_{u=0}$.
Hence, in general this theory described 6 degrees of freedom, as we found 
it already in linearized calculations on the selfaccelerated background. 
For appropriate choice of boundary conditions these could be 
a massive graviton plus an additional scalar,
which have no ghosts or tachyons, as it was shown in 
Section 3. For some particular boundary conditions though, 
due to enhanced symmetries of the linearized perturbations, 
the number of linear degrees of freedom   gets reduced. 
In this case some of the equations in (\ref {N2}) and (\ref {Nj2}) 
should appear as constraints in the linearized theory.

\subsection*{5. Discussions}

The extension of GR considered in this work is a convenient 
way of putting  various  theories of massive gravity 
in a single framework. All these theories, 
known in the linearized  level, emerge as a consequence of 
choosing different boundary conditions in the  auxiliary dimension.
Moreover, the present framework provides a non-linear completion to these
theories with the Hamiltonian that does not suffer from the 
problems found in Ref. \cite {Deser}. The auxiliary dimension is 
just a convenient technical  tool; it  can in principle be 
``integrated out'' entirely, and  this should lead to 
GR amended by new terms in 4D.

Most importantly, the extended theory admits the 
selfaccelerated solution with the spectrum of linear perturbations 
that has no ghosts or tachyons. In a general case one obtains 
massive graviton and a scalar. This may have some cosmological 
signatures  along the lines of Refs. \cite {Justin,Roman}.   
The vDVZ discontinuity of the linearized theory has to be 
overcome through the strong dynamics via the 
Vainshtein mechanism \cite {Arkady} (see also \cite {DDGV}). 
If this is the case, then  the theory is likely to have also short 
distance signatures \cite {Lue1}, \cite {DGZ}.
  
We end this section by a few comments.

The auxiliary dimension discussed so far had a finite extent in the 
$u$-direction. It is straightforward to present a Lagrangian 
in which the $u$-direction is infinite:
\beq
\sqrt{g} R +  {m^2}\int_{-\infty }^{+\infty}du \sqrt{\t g}
\left (  k_{\mu\nu}^2 - k^2 -3 \right )\,.
\label{egrinfty}
\eeq
The equations of motion of this Lagrangian have 
a selfaccelerated solution $\tg_{\mu\nu}=a(u)\bga (x)$, 
where $a(u) =e^{-u}$, and as before, $\bga$ denotes the 4D de Sitter metric
with curvature $R=12m^2$.

The Lagrangians (\ref {egr}) and  (\ref {egrinfty}), 
can be obtained by a certain  truncation of  
a 5D theory. The 5D theory giving   (\ref {egrinfty})
can be defined as  follows:    
\beq
\sqrt{g} R +  m_c\int dy  \sqrt{g^{(5)}} \left ( R(\t g) -R_5(g^{(5)}) - 
3{\t m}^2_c 
\right )|_{g_{55}=1,g_{\mu 5}=0 }\,,
\label{5D}
\eeq
where $R_5$ is the 5D Ricci scalar, $g^{(5)}_{AB}= \{ \tg_{\mu \nu}, 
g_{\mu 5}, g_{55} \},~~A,B=0,1,2,3,5$, 
$y= u/{\t m}_c$, $m^2=m_c {\t m}_c$, and the substitutions  in 
the last term are 
taken before the equations of motion are obtained, i.e., there is no 
variation w.r.t. $ g_{55} $  and  $g_{\mu 5}$. To get the analogous 
expression  for (\ref {egr}) one would have to drop the last term in 
the parenthesis, and  set the integration w.r.t. $y$ from 
$ -1/ {\t m}_c $ to $ +1/{\t m}_c $. The expression (\ref {5D})  
is somewhat similar to the DGP Lagrangian  \cite {DGP}, or 
its sign-flipped counterpart \cite {GGCargese}, with two  
crucial differences: (1) There is a subtraction of the $R$ 
term from the $R_5$ term in the bulk action; (2) There are no $\{55\}$ or 
$\{\mu 5\}$ equations\footnote{One gets back the analog of the 
$\{\mu 5\}$ equation in Eq. (\ref {Con}) due to the Bianchi identities.
It is really the absence of the $\{55\}$ equation that makes a crucial 
difference.}.

Similar constructions with an auxiliary dimension can be 
considered for a scalar or vector, by adding the term 
$-m^2 \int du [(\partial_u\phi)^2 + \phi^2+...]$, to the conventional scalar 
field Lagrangian,  or the term  $-m^2 \int du [(\partial_u A_\mu)^2 +...]$
to the  Maxwell Lagrangian (with a finite or an infinite range of integration).

\subsection*{Acknowledgments}

I'd like to thank Massimo Porrati for useful discussions.
The work was partially  supported by the NSF grant PHY-0758032.

\end{document}